\documentclass[12pt]{iopart}
\usepackage[dvips]{graphicx}
\begin{document}
\title[First-Order Phase transition in a Reaction-Diffusion Model]
{First-Order Phase Transition in a Reaction-Diffusion Model With
Open Boundary: The Yang-Lee Theory Approach}
\author{Farhad H Jafarpour
\footnote[1]{e-mail:farhad@sadaf.basu.ac.ir}}
\address{Bu-Ali Sina University, Physics Department, Hamadan, Iran}
\begin{abstract}
A coagulation-decoagulation model is introduced on a chain of
length $L$ with open boundary. The model consists of one species
of particles which diffuse, coagulate and decoagulate
preferentially in the leftward direction. They are also injected
and extracted from the left boundary with different rates. We will
show that on a specific plane in the space of parameters, the
steady state weights can be calculated exactly using a matrix
product method. The model exhibits a first-order phase transition
between a low-density and a high-density phase. The density
profile of the particles in each phase is obtained both
analytically and using the Monte Carlo Simulation. The two-point
density-density correlation function in each phase has also been
calculated. By applying the Yang-Lee theory we can predict the
same phase diagram for the model. This model is further evidence
for the applicability of the Yang-Lee theory in the
non-equilibrium statistical mechanics context.
\end{abstract}
\submitto{\JPA}
\maketitle
\section{Introduction}
It is known that one-dimensional driven diffusive systems can
exhibit many interesting collective phenomena such as jamming and
spontaneous symmetry breaking \cite{sz,sch}. Applications of such
models are divers and include the kinetics of biopolymerization
\cite{MGP} and traffic flow \cite{CSS}, but to give just a few
examples. \\
A powerful framework for studying equilibrium phase
transitions in classical statistical mechanics is the Yang-Lee
theory of phase transition \cite{yanglee1,yanglee2}. Here we shall
briefly explain the key points of this theory. Consider a simple
model consisting of $L$ spins in thermal contact with a heat
reservoir. The hallmark of a phase transition in this system is
the appearance of nonanalyticities in its free energy which can be
defined in terms of the partition function of the system $Z_{L}$
as $f=\frac{1}{L}\ln Z_{L}$. In the thermal equilibrium the
partition function (which is a function of temperature $T$ in our
example) determines the statistical properties of the system.
However, since the partition function has no real and positive
zeros in the complex-$T$ plane, there is no scope for a phase
transition in a finite system. As we increase the number of spins
$L$ to infinity, the partition function zeros might accumulate
towards a point $T_0$ on the real axis so there is the possibility
of a phase transition at this point. The density of zeros near the
accumulation point determines the order of the phase transition.
At a first-order phase transition there is a nonzero density of
zeros at the critical point whereas the density of zeros decays to
zero at a continuous transition. Although we explained the theory
of partition function zeros with reference to a system described
by a canonical partition function, we should note that the theory
still holds when one is working in grand canonical ensembles. In
this case the fugacity (or chemical potential) plays similar role
as the temperature $T$. More generally one can look for the zeros
of the partition function in the complex plane of any of its
intensive variables
\cite{gross1,gross2}.\\
The lack of such a theory in the non-equilibrium statistical
mechanics context has recently encouraged people to examine
whether a generalized version of it can still work for
out-of-equilibrium systems. Quite similar to the equilibrium
statistical mechanics one can define a generalized partition
function as sum over the steady state weights. These weights can
be calculated using different methods such as matrix product
formalism in which the stationary probability distribution
function is written in terms of the products of non-commuting
operators \cite{dehp}. Then we look at the zeros of the partition
function in the thermodynamic limit in the complex plane of any
intensive variable of the model. The zeros might divide the
complex plane into several regions where the free energy of the
system is analytic. Each region corresponds to a different phase
and as we explained above, the order of transition depends on how
the zeros accumulate towards the positive real axis. Several out
of equilibrium models have been proposed and studied using the
generalized version of the Yang-Lee theory
\cite{farjaf}-\cite{hin}, and it seems that the results obtained
from application of this theory accord themselves well
with the known results obtained from other approaches.\\
In this paper we will introduce a one-dimensional
reaction-diffusion model with open boundary and apply the Yang-Lee
theory thereto. This model has already been studied with
reflecting boundaries \cite{HKP1,HKP2} while the open boundary
problem has not been solved. We shall investigate the effects of
the open boundary condition on the phase diagram and also the
phase transition points of this model. Our work will also be
another example (with more complicated reaction rules) that shows
the applicability of the Yang-Lee theory to study the critical
behaviors of the out of equilibrium systems. \\
In chapter (2) we define the model. The chapter (3) is devoted to
the application of the matrix product formalism  in order to find
the steady states weights and therefore the partition function of
the model. Using this formalism allows us to find the stationary
weights of the model exactly and calculate many interesting
quantities such as the density profile of particles on the chain
and also the correlation functions. It turns out that our model
has a first-order phase transition between a low-density and a
high-density phase provided that there is a constraint on the
reaction rates. In chapter (4) we shall apply the Yang-Lee theory
to our model and calculate the grand canonical partition function
of the system and its line of zeros in the thermodynamic limit.
The line of zeros is a circle and the density of zeros is constant
over it all. In the Yang-Lee theory language this signifies a
first-order phase transition. Finally in section (5) we shall
conclude and explain the generalizations.
\section{The Model}
Consider a one-dimensional lattice of length $L$. From the left
boundary classical particles are injected with rate $\alpha$ if
the target site is empty. They can also be extracted from there
with rate $\beta$, provided that it is occupied by a particle. In
the bulk of the chain particles diffuse to the left and right.
When two of them meet, they can merge into a single particle.
Similarly, a single particle can split into two particles. All of
these processes take place preferentially to the leftward
direction. Specifically, the reaction rules in the bulk of the
chain are
\begin{equation}
\label{P1}
\begin{array}{llll}
\mbox{diffusion to the left:} && \emptyset+A \rightarrow
A+\emptyset &
\mbox{with rate} \; \; q \\
\mbox{diffusion to the right:} && A+\emptyset \rightarrow
\emptyset+A &
\mbox{with rate} \; \; q^{-1} \\
\mbox{coagulation at the left:} && A+A \rightarrow A+\emptyset &
\mbox{with rate} \; \; q \\
\mbox{coagulation at the right:} && A+A \rightarrow \emptyset+A &
\mbox{with rate} \; \; q^{-1} \\
\mbox{decoagulation to the left:} && \emptyset+A \rightarrow A+A &
\mbox{with rate} \; \; \Delta q \\
\mbox{decoagulation to the right:} && A+\emptyset \rightarrow A+A
& \mbox{with rate} \; \; \Delta q^{-1}
\end{array}
\end{equation}
in which $A$ and $\emptyset$ represent an occupied and an empty
site respectively. At the left boundary we have
\begin{equation}
\label{P2}
\begin{array}{llll}
\mbox{injection at the first site:} & & \emptyset \rightarrow A &
\mbox{with rate} \; \; \alpha\\
\mbox{extraction at the first site:} & & A \rightarrow \emptyset &
\mbox{with rate} \; \; \beta.\\
\end{array}
\end{equation}
It is seen that (\ref{P1}) consists of transitions which involve
only the configurations of nearest neighboring sites. For $q>1$
the particles have a tendency to move in the leftward direction.
In the following section we will use the Matrix Product Formalism
to find the steady states weights of this model.
\section{Application of the Matrix Product Formalism}
The time evolution operator of our model can be written as
$$
H=\sum_{j=1}^{L-1} h_{j,j+1} + h^{(L)}_1
$$
in which
$$
h_{j,j+1}={\cal I}^{\otimes (i-1)}\otimes h \otimes {\cal
I}^{\otimes (L-i-1)}
$$
and
$$
h^{(L)}_1=h^{(L)} \otimes {\cal I}^{\otimes (L-1)}
$$
where ${\cal I}$ is a $2 \times 2$ identity matrix, $h$ is a $4
\times 4$ matrix for the bulk interactions and $h^{(L)}$ is a $2
\times 2$ matrix for particle input and output at the left
boundary. In a basis $(\emptyset\emptyset,\emptyset
A,A\emptyset,AA)$ the bulk evolution operator $h$ and $h^{(L)}$
read
\begin{equation}
\label{Hamiltonian}
h=\left(
\begin{array}{cccc}
0 & 0 & 0 & 0 \\
0 & -(\Delta+1) q & q^{-1} & q^{-1} \\
0 & q & -(\Delta+1) q^{-1} & q \\
0 & \Delta q & \Delta q^{-1} & -q+q^{-1}
\end{array} \right) ,
h^{(L)}=\left(
\begin{array}{cc}
-\alpha & \beta  \\
\alpha & -\beta
\end{array} \right).
\end{equation}
In the following we show that the stationary probability
distribution of any configuration of our model can be calculated
exactly using the matrix product formalism (MPF)\cite{dehp}. Let
us first briefly review the MPF. According to this approach, for
models with Hamiltonian $H=\sum_{j=1}^{L-1} h_{j,j+1} +
h^{(L)}_1+h^{(R)}_L$, the stationary probability distribution
$P({\cal C})$ of any configuration ${\cal C}$ is assumed to be of
the form
\begin{equation}
\label{Weigth} P({\cal C})=\frac{1}{Z_{L}} \langle W \vert
\prod_{i=1}^{L}(\tau_i D+(1-\tau_i)E)\vert V \rangle
\end{equation}
($\tau_i=0$ if the site $i$ is empty and $\tau_i=1$ if it is
occupied by a particle) with the following property
$$
H P({\cal C})=0.
$$
The factor $Z_L$ in (\ref{Weigth}) is a normalization factor and
can easily be obtained using the normalization condition
$\sum_{\cal C}P({\cal C})=1$. The matrices $D$ and $E$ are square
matrices and stand for the presence of a particle and an empty
site. These matrices beside the vectors $\langle W \vert$ and
$\vert V \rangle$ satisfy the following algebra
\begin{equation}
\begin{array}{l}
\label{MPA}
h  \left[ \left( \begin{array}{c} E \\
D \end{array} \right) \otimes
\left( \begin{array}{c} E \\
D \end{array} \right) \right]=
\left( \begin{array}{c} \bar{E} \\
 \bar{D} \end{array} \right) \otimes
\left( \begin{array}{c} E \\
 D \end{array} \right) -
\left( \begin{array}{c} E \\
 D \end{array} \right) \otimes
\left( \begin{array}{c} \bar{E} \\
 \bar{D} \end{array} \right), \\ \\
<W| h^{(L)} \left( \begin{array}{c} E \\
 D \end{array} \right) =
-<W| \left( \begin{array}{c} \bar{E} \\
 \bar{D} \end{array} \right), \\ \\
h^{(R)} \left( \begin{array}{c} E \\
 D \end{array} \right) |V> =
\left(\begin{array}{c} \bar{E} \\
 \bar{D} \end{array} \right) |V>. \nonumber
\end{array}
\end{equation}
The auxiliary matrices $\bar{E}$ and $\bar{D}$ are also square
matrices. For the present model with $h^{(R)}=0$, one can easily
see from (\ref{Hamiltonian}) and (\ref{MPA}) that the
corresponding algebra is
\begin{equation}
\begin{array}{l}
\label{FinalBulkAlgebra}
[C,\bar{C}] = [E,\bar{E}] = 0 \\ \\
\bar{E}C-E\bar{C} =(q+q \Delta+ q^{-1}) EC - q(1+\Delta)  E^{2} -
q^{-1} C^{2} \\ \\
\bar{C}E-C\bar{E} =(q^{-1}+q^{-1} \Delta+ q)  CE -q^{-1}(1+\Delta)
E^{2} - q C^{2}.
\end{array}
\end{equation}
with
\begin{equation}
\begin{array}{l}
\label{FinalSurfaceAlgebra} \langle W
\vert((\alpha+\beta)E+\bar{E}-\beta
C)=\langle W \vert \bar {C}=0 \\ \\
\bar{E} \vert V \rangle =\bar{C}\vert V \rangle=0
\end{array}
\end{equation}
in which we have defined $C:=D+E$ and $\bar{C}:=\bar{D}+\bar{E}$.
Traditionally, one should find either a representation for the
algebra (\ref{FinalBulkAlgebra}) and (\ref{FinalSurfaceAlgebra})/
or calculate the stationary weights (\ref{Weigth}) directly from
the algebra without using any representations. In the following we
will show that a finite-dimensional representation of the algebra
(\ref{FinalBulkAlgebra}) and (\ref{FinalSurfaceAlgebra}) exists
under special constraint on the parameters $\alpha$, $\beta$, $q$
and $\Delta$. One can easily check that for $q^{2}\neq1+\Delta$
the following matrices and vectors
\begin{equation}
\begin{array}{l}
\label{RepBulk1}
C=\left(\begin{array}{cc}
1+\Delta & 0  \\
0 & q^{2}
\end{array} \right) ,\bar{C}=0,
E=\left(\begin{array}{cc}
1 & \lambda  \\
0 & q^{2}
\end{array} \right),
\bar{E}=\left(\begin{array}{cc}
\frac{q^{2}-1}{q}\Delta & -\frac{\Delta}{q}\lambda  \\
0 & 0
\end{array} \right)\\ \\
\vert V \rangle=\left(\begin{array}{c} \frac{\lambda}{q^{2}-1}\\1
\end{array} \right),\langle W \vert=\Bigl(\frac{q\Delta(q^{2}-q
\beta-1)}{\lambda(\beta+\beta\Delta-q\Delta)},1 \Bigr)
\end{array}
\end{equation}
provide a two-dimensional representation for the algebra
(\ref{FinalBulkAlgebra}) and (\ref{FinalSurfaceAlgebra}) provided
that $\alpha=(q^{-1}-q+\beta)\Delta$. For all rates to be positive
we assume $q > 1$ and $\beta\geq q-q^{-1}$. For $q^{2} =
1+\Delta$, $C$ is not diagonalizable and we find another
representation
\begin{equation}
\begin{array}{l}
\label{RepBulk2}
C=\left(\begin{array}{cc}
1 & \lambda  \\
0 & 1
\end{array} \right) ,\bar{C}=0,
E=\left(\begin{array}{cc}
\frac{1}{q^{2}} & \lambda \\
0 & 1
\end{array} \right),
\bar{E}=\left(\begin{array}{cc}
\frac{(q^{2}-1)^{2}}{q^{3}}& -\frac{q^{2}-1}{q}\lambda  \\
0 & 0
\end{array} \right)\\  \\
\vert V \rangle=\left(\begin{array}{c}
\frac{q^{2}\lambda}{q^{2}-1}\\1
\end{array} \right),\langle W \vert=\Bigl(-\frac{q^{3}-\beta q^{2}-2q+q^{-1}+\beta}
{\lambda(q^{3}-\beta q^{2}-q+\beta)},1 \Bigr).
\end{array}
\end{equation}
In (\ref{RepBulk1}) and (\ref{RepBulk2}), $\lambda$ is a free
parameters. Using the representations (\ref{RepBulk1}) and
(\ref{RepBulk2}) one can easily calculate the density profile of
the particles in the stationary state and also their
density-density correlations. The mean density of particle on the
chain at site $i$, $\rho_i$, can be written in terms of the
matrices $C$ and $E$ and the vectors $\vert V \rangle$ and
$\langle W \vert$
\begin{equation}
\label{DensityProfile} \langle \rho_i \rangle=\frac{\langle W
\vert C^{i-1}(C-E)C^{L-i}\vert V \rangle}{\langle W \vert C^L
\vert V \rangle}.
\end{equation}
By using (\ref{RepBulk1}) for $q^{2}\neq 1+\Delta$ and after some
algebra we find
\begin{eqnarray}
\label{DP1}
\langle \rho_i \rangle=  \\
\frac{\Delta(1+\Delta)^{-1}(1-q^2+q\beta)(\Delta(1+\Delta)^L
q^{2i}-(q^2-1)(1+\Delta)^iq^{2L})q^{-2i+1}}{q\Delta
(q^2-1)(q^{2L}-(1+\Delta)^L)+\beta(\Delta q^2
(1+\Delta)^L-q^{2L}(q^2-1)(1+\Delta))}. \nonumber
\end{eqnarray}
Two different phases are apparent: $q^2 > 1+\Delta$ and $q^2 <
1+\Delta$. In the thermodynamic limit $L\rightarrow \infty$ we
find two different expressions for the mean density of particles
at site $i$. In the first phase, which is called the
\emph{low-density phase}, we obtain
\begin{equation}
\label{LDensity}\langle \rho_i \rangle= \frac{q^2
\Delta(q^{-1}-q+\beta)}{(1+\Delta)(\beta(1+\Delta)-q\Delta)}
e^{-i/\xi}
\;\;\;\;\mbox{for} \;\;\;\; q^2>1+\Delta.
\end{equation}
As can be seen, the density of particles at the left boundary has
the largest value and will quickly drop to zero in the bulk of the
chain with characteristic length $\xi$ where $\xi^{-1}=\vert \ln
\frac{q^2}{1+\Delta}\vert$.
\begin{figure}[htbp]
\setlength{\unitlength}{1mm}
\begin{picture}(0,0)
\put(-5,27){\makebox{$\scriptstyle \langle\rho_i\rangle$}}
\put(42,-2){\makebox{$\scriptstyle i$}}
\end{picture}
\centering
\includegraphics[height=5cm] {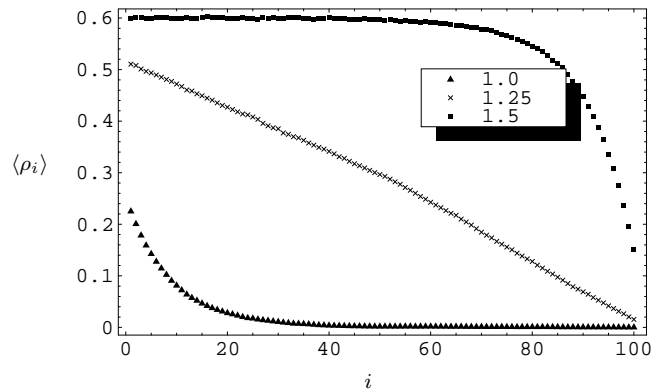}
\caption{Monte Carlo simulation results for the density profile of
the particles on a chain of length $L=100$ for $q=1.5$,
$\beta=0.9$ and different values of $\Delta$. The input rate
$\alpha$ is given by $\alpha=(q^{-1}-q+\beta)\Delta$. }
\end{figure}
For the second phase, the \emph{high-density phase}, we obtain
\begin{equation}
\label{HDensity} \langle \rho_i
\rangle=\frac{\Delta}{1+\Delta}-\frac{q^2-1}{1+\Delta}
e^{-(L-i)/\xi}
\;\;\;\;\mbox{for} \;\;\;\; q^2<1+\Delta
\end{equation}
where $\xi^{-1}=\vert \ln \frac{q^2}{1+\Delta}\vert$. In
high-density phase the density of particles is nearly constant in
the bulk of the chain equal to $\frac{\Delta}{1+\Delta}$ and drops
to zero near the right boundary. For $q^2=1+\Delta$ we use
(\ref{RepBulk2}) to find the following expression for the mean
density of particles at site $i$ in the thermodynamic limit
\begin{equation}
\label{Shock} \langle \rho(x)
\rangle=\frac{\Delta}{1+\Delta}(1-x)\;\;\;\;\mbox{for} \;\;\;\;
q^2=1+\Delta
\end{equation}
in which $x:=\frac{i}{L}$ and $0 \leq x\leq 1$. In Figure (1) we
have plotted the density profile of the particles on a chain of
length $L=100$ for different values of $\Delta$ obtained from the
Monte carlo simulation. One can readily check that the Monte Carlo
results accord with the exact analytical results
(\ref{LDensity})-(\ref{Shock}). In order to see the phase
transition more clearly let us define the density of particles
averaged over the entire system as an order parameter
\begin{equation}
\label{AverageDensity} \rho:=\frac{1}{L}\sum_{i=1}^L\langle \rho_i
\rangle.
\end{equation}
Plot of (\ref{AverageDensity}) as a function of $\Delta$ for
$q=3$, $\beta=3$ and two different values of system length
$L=200,2000$ is shown in the Figure (2). It can be seen that as
the length of the chain approaches the thermodynamic limit
$L\rightarrow \infty$, a discontinuity appears in $\rho$ at the
transition point $\Delta_c=q^2-1$ which illustrates a first-order
phase transition in the system.
\begin{figure}[htbp]
\setlength{\unitlength}{1mm}
\begin{picture}(0,0)
\put(-4,26){\makebox{$\scriptstyle \rho$}}
\put(45,-1){\makebox{$\scriptstyle \Delta$}}
\end{picture}
\centering
\includegraphics[height=5cm] {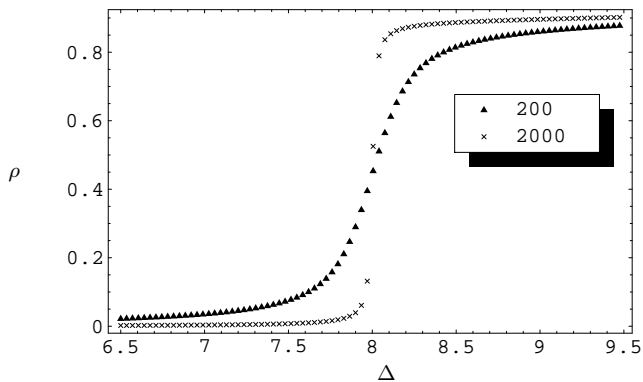}
\caption{Plot of (\ref{AverageDensity}) as a function of $\Delta$
for systems of length $L=200$ and $L=2000$ with $q=3$ and
$\beta=3$. In the thermodynamic limit $L\rightarrow\infty$ a
first-order phase transition occurs at $\Delta_c=q^2-1=8$.}
\end{figure}
Using the matrix representation of the algebra
(\ref{FinalBulkAlgebra}) and (\ref{FinalSurfaceAlgebra}) we can
also calculate two-point density correlation functions of this
model in each phase exactly. The connected two-point
density-density correlation function is defined as
\begin{equation}
\label{CTPCF} \langle \rho_i\rho_j\rangle_c=\langle
\rho_i\rho_j\rangle-\langle\rho_i\rangle\langle\rho_j\rangle \;\;
\mbox{for} \;\; i<j.
\end{equation}
As we saw in (\ref{DensityProfile}), the density of particles on
the chain at an arbitrary site $i$ can be written in terms of the
operators $C$ and $E$ and the boundary vectors $\langle W \vert$
and $\vert V\rangle$. The expression $\langle \rho_i\rho_j\rangle$
can also be written in terms of the same operators and the vectors
as
\begin{equation}
\label{TPCF1} \langle \rho_i\rho_j\rangle=\frac{\langle W \vert
C^{i-1}(C-E)C^{j-i-1}(C-E)C^{L-j}\vert V \rangle}{\langle W \vert
C^L \vert V \rangle}.
\end{equation}
By using the matrix representation of the algebra we find
\begin{equation}
\label{TPCF2} \langle \rho_i\rho_j\rangle=
\frac{q{\Delta^2}(q^{2}-q\beta-1)}{Z_{L}(\beta+\beta\Delta-\Delta
q)}[\frac{\Delta(1+\Delta)^{L-2}}{q^{2}-1}-
q^{2(L-j)}(1+\Delta^{j-2})]
\end{equation}
in which $Z_{L}=\langle W \vert C^L \vert V \rangle$ is given by
(\ref{PF}). Now we look for the thermodynamic limit of
(\ref{CTPCF}) in each phase. Using (\ref{DP1}), (\ref{CTPCF}) and
(\ref{TPCF2}) we obtain the following expressions for the
thermodynamic limit of the connected two-point correlation
function of particles
\begin{equation}
\label{CTPCF2}
\langle\rho_i\rho_j\rangle_c=(\frac{\Delta}{1+\Delta}-\langle
\rho_i \rangle)\langle \rho_j \rangle.
\end{equation}
One should use (\ref{LDensity}) or (\ref{HDensity}) for $\langle
\rho_i \rangle$ in each phase. This expression is valid for both
$q^2>1+\Delta$ and $q^2<1+\Delta$ and has also a simple
explanation; namely, in the low-density phase the connected
two-point correlation function (\ref{CTPCF}) is non-zero only in
the close vicinity of the left boundary where $i < j \sim \xi$.
For $j\gg \xi$, $\langle\rho_i\rho_j\rangle_c\equiv 0$. Thus, in
the whole region $\xi \ll j < L$, particles do not feel any
correlations. In the high-density phase both $i$ and $j$ in
(\ref{CTPCF}) should be near the last site of the chain far from
the left boundary, otherwise $\langle\rho_i\rho_j\rangle_c\equiv
0$.
\section{Application of the Yang-Lee Theory}
Now we will examine the Yang-Lee theory. We define the
partition function of the system $Z_{L}$ as the sum of the
stationary states weights given by (\ref{Weigth})
$$Z_{L}=\sum_{\cal C} P({\cal C})=\sum_{ \{\tau_i=0,1\} }\langle W
\vert \prod_{i=1}^{L}(\tau_i D+(1-\tau_i)E)\vert V \rangle=\langle
W \vert C^{L} \vert V\rangle.$$
Using the matrix representation of the operator $C$ and the
vectors $\vert V \rangle$ and $\langle W \vert$ in
(\ref{RepBulk1}) one obtains
\begin{equation}
\label{PF} Z_L=q^{2L}+\frac{\Delta q (q^2-q
\beta-1)(1+\Delta)^L}{(q^2-1)(\beta+\beta \Delta-q \Delta)}.
\end{equation}
We look for the zeros of our partition function (\ref{PF}) in the
complex-$q$ plane at fixed $\Delta$ and $\beta$ in the
thermodynamic limit $L\rightarrow \infty$. In Figure (3) we have
plotted the numerical estimates of these roots for $\Delta=8$,
$\beta=3$ and $L=100$. It can be seen that the roots lie on a
circle of radius $q_c=\sqrt{1+\Delta}=3$, as we had predicted
above. The roots also intersect the real-$q$ axis at an angle
$\frac{\pi}{2}$ which is the sign of first-order phase transition.
\begin{figure}[htbp]
\setlength{\unitlength}{1mm}
\begin{picture}(0,0)
\put(-10,27){\makebox{$\scriptstyle Im \;\; q$}}
\put(23,-2){\makebox{$\scriptstyle Re \;\; q$}}
\end{picture}
\centering
\includegraphics[height=5cm] {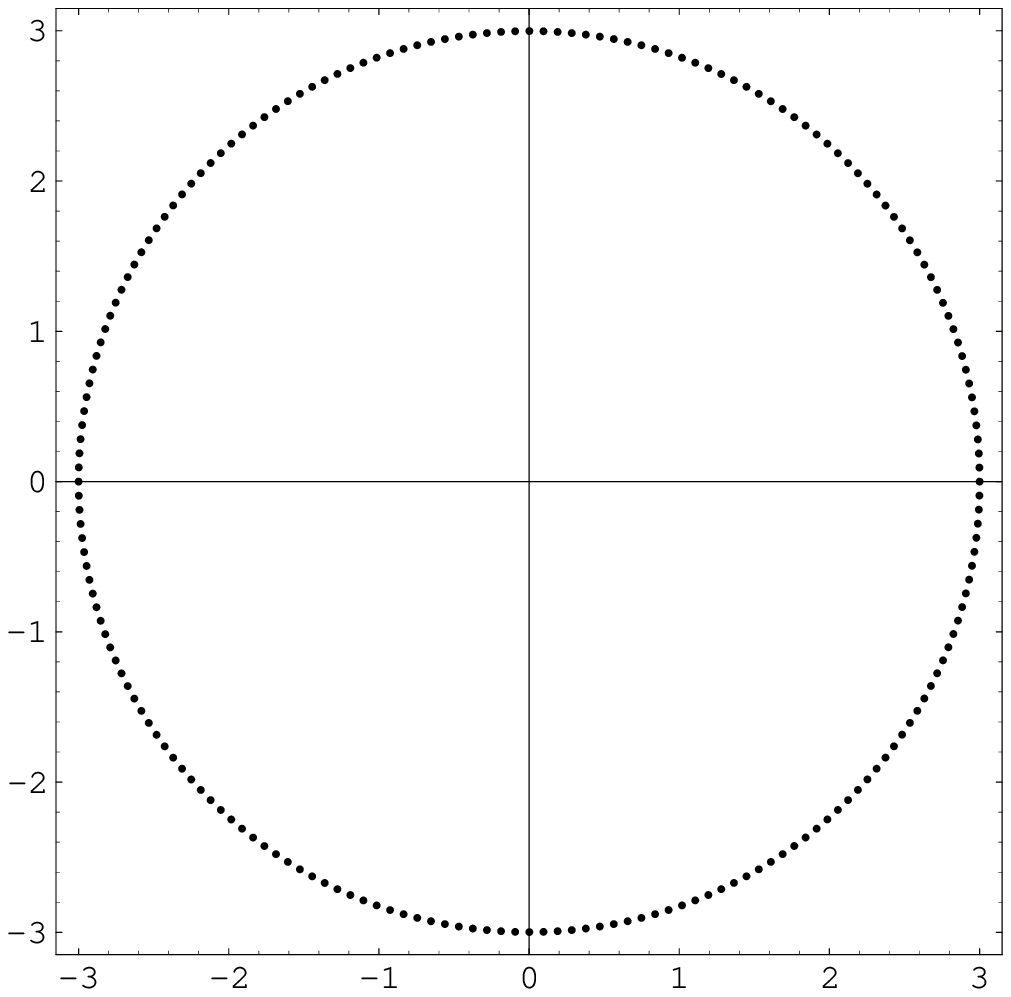}
\caption{Plot of the numerical estimates for the roots of
(\ref{PF}) in the complex $q$ plane for $L=100$, $\Delta=8$ and
$\beta=3$.}
\end{figure}
Beside the numerical estimates, one can obtain the line of zeros
and also their density near the positive real $q$ axis
analytically. Following \cite{gross1,gross2} we define the
extensive part of the free energy as
\begin{equation}
\label{FreeEnergy} g(q,\Delta,\beta)=\lim_{L\rightarrow
\infty}\frac{1}{L} \ln Z_{L}
\end{equation}
in which $Z_{L}$ is given by (\ref{PF}). Now the line of zeros can
be obtained from
\begin{equation}
\label{LineEqu} Re\; g_1=Re\; g_2.
\end{equation}
Here $g_1$ and $g_2$ are the thermodynamic limits of $g$ on the
left and the right hand side of the phase transition point. By
using (\ref{PF}), (\ref{FreeEnergy}) and (\ref{LineEqu}) and after
some calculations one finds the line of zeros
\begin{equation}
\label{Circle} x^2+y^2=(1+\Delta).
\end{equation}
In (\ref{Circle}) the parameters $x$ and $y$ are defined as the
real and the imaginary parts of $q$ respectively. This result is
in quite agreement with the numerical estimate of zeros given in
Figure (3). In order to find the density of zeros $\mu$ near the
positive real $q$ axis, we use \cite{gross1,gross2}
\begin{equation}
\label{RootsDensity} 2 \pi
\mu(s,\Delta,\beta)=\frac{\partial}{\partial s} Im(g_{1}-g_{2})
\end{equation}
where $s$ is the arc length of the line of zeros which is measured
from the critical point and increases along with the line of
zeros. The functions $g_1$ and $g_2$ have the same definitions as
mentioned above. After some algebra we find
$\mu=\frac{1}{\pi\sqrt{1+\Delta}}$. It is seen that the density of
zeros is constant everywhere on the circle beside in the vicinity
of the positive real $q$ axis; therefore, it confirms the
existence of a first-order phase transition in our model.
\section{Concluding Remarks}
In this paper we introduced a one-dimensional out-of-equilibrium
model consisting of one class of particles which diffuse,
coagulate and decoagulate on an open chain. The left boundary of
the chain is assumed to be open so that the particles can enter or
leave the system. Exact calculations using the MPF show that on a
specific plane in the parameters space a first-order phase
transition takes place between a low-density and a high-density
phase. The density profile of the particles in each phase is
obtained exactly. In the low-density phase the density profile of
particles far from the left boundary drops exponentially to zero
with the length scale $\xi=\vert\ln
\frac{q^2}{1+\Delta}\vert^{-1}$. We should mention that in the
stationary state there exists only one characteristic length scale
$\xi$, while the same model without injection and extraction of
particles has three different length scales \cite{HKP1,HKP2}. In
the high-density phase the density of particles is nearly constant
in the bulk; however, it drops to zero near the right boundary
with the same length scale $\xi$. It is shown that the steady
state of our model can be written in terms of the superposition of
shocks \cite{JKS}. In order to answer the question whether we can
apply the classical Yang-Lee theory to the out-of-equilibrium
systems such as the one introduced here, we have seen that the
application of this theory gives same results obtained from the
MPF approach. We calculated the line of partition function zeros
exactly. It intersects the positive real $q$ axis at right angles
at $q_c=\sqrt{1+\Delta}$. The density of zeros is uniform at the
cross point, which means a first-order phase transition at
critical point. Our model, which is more complex than those
studied before, is another example for the generality of the
classical Yang-Lee picture of phase transitions in one-dimensional
out-of-equilibrium models. It can be shown that the MPF also works
when the particles are allowed to leave the chain from the right
boundary; however, there should be still a constraint on the
reaction rates. The whole phase diagram of this model which can be
studied at the mean field level or using the Monte-Carlo
simulation, is still an open problem.

\end{document}